\documentclass[11pt]{article}
\usepackage{latexsym}  
\usepackage{amssymb}
\usepackage{graphicx}
\usepackage{amsmath}

\begin{document}

\begin{center}
{\Large\bf Testable Deviations from Exact Tribimaximal Mixing}

\vspace{4mm}
{\bf Yoshio Koide$^a$ and Hiroyuki Nishiura$^b$}

${}^a$ {\it IHERP, Osaka University, 1-16 Machikaneyama, 
Toyonaka, Osaka 560-0043, Japan} \\
{\it E-mail address: koide@het.phys.sci.osaka-u.ac.jp}

${}^b$ {\it Faculty of Information Science and Technology, 
Osaka Institute of Technology, 
Hirakata, Osaka 573-0196, Japan}\\
{\it E-mail address: nishiura@is.oit.ac.jp}

\date{\today}
\end{center}

\vspace{3mm}
\begin{abstract}
A simple relation $U_{PMNS}=V_{CKM}^\dagger U_{TB}$
between the lepton and quark mixing matrices ($U_{PMNS}$ and
$V_{CKM}$) is
speculated under an ansatz that $U_{PMNS}$ becomes an exact 
tribimaximal mixing $U_{TB}$ in a limit $V_{CKM}={\bf 1}$.  
By using the observed CKM mixing parameters, possible values 
of neutrino oscillation parameters are estimated: 
$\sin^2 \theta_{13}=0.024-0.028$,
$\sin^2 2\theta_{23}=0.94-0.95$ and 
$\tan^2 \theta_{12}=0.24-1.00$  depending  
on phase conventions of $U_{TB}$.
Those values are testable soon by precision measurements 
in neutrino oscillation experiments.
\end{abstract}

\vspace{3mm}

\noindent{\large\bf 1 \ Introduction}

Recently, there has been considerable interest in the magnitude
of the neutrino mixing angle $\theta_{13}$ 
($\nu_e \leftrightarrow \nu_\tau$ mixing angle), because it is a key 
value  not only for checking neutrino mass matrix models, but
also for searching $CP$-violation effects in the lepton sector.
(For a review of models for $\theta_{13}$, see, for example, 
Ref.\cite{U13}.)
Recent observed neutrino oscillation data  are in favor of the 
so-called ``tribimaximal mixing" \cite{tribi} which predicts
$\theta_{13}=0$, $\tan^2\theta_{12}=1/2$ and $\sin^2 2\theta_{23}=1$,
since the present data yield the values 
$\tan^2 \theta_{12} = 0.47^{+0.06}_{-0.05}$ \cite{KamLAND} and
$\sin^2 2\theta_{23}=1.00_{-0.13}$ \cite{MINOS}.
If the angle $\theta_{13}$ is exactly zero or negligibly small, 
the observation of the $CP$-violation effects in the lepton sector 
will be hopeless  even in future, as far as neutrino oscillation
experiments are concerned.
On the other hand, recently, Fogli {\it et al.} \cite{Fogli08}
have reported a sizable value $\sin^2\theta_{13}=0.016\pm 0.010$
($1\sigma$) from a global analysis of neutrino oscillation data.

The tribimaximal lepton mixing is given by the form
$$
U_{TB}^0 = \left( 
\begin{array}{ccc}
\frac{2}{\sqrt6} & \frac{1}{\sqrt3} & 0 \\
-\frac{1}{\sqrt6} & \frac{1}{\sqrt3} & -\frac{1}{\sqrt2} \\
-\frac{1}{\sqrt6} & \frac{1}{\sqrt3} & \frac{1}{\sqrt2}
\end{array} \right) .
\eqno(1)
$$
Such a form with beautiful coefficients seems to be understood
from a discrete symmetry of flavors \cite{tribi}. 
In contrast to the lepton mixing matrix 
(Pontecorvo-Maki-Nakagawa-Sakata mixing matrix \cite{MNS}) 
$U_{PMNS}$, the observed Cabibbo-Kobayashi-Maskawa \cite{CKM} 
(CKM) quark mixing matrix $V_{CKM}$ seems to have no 
beautiful form with Clebsch-Gordan-like coefficients, and 
$V_{CKM}$, rather, looks like nearly $V_{CKM}\simeq {\bf 1}$.
It is unlikely that a theory which exactly leads to the tribimaximal 
mixing (1) simultaneously gives the CKM mixing matrix with 
small and complicated mixing values. 
Therefore, it is interesting to consider a specific case that 
a theory of flavor symmetry gives $V_{CKM}={\bf 1}$ 
in the limit of $U_{PMNS}=U_{TB}$.  
We consider that the observed form of the CKM matrix $V_{CKM}$ is 
due to some additional effects (e.g. symmetry breaking effects 
for the flavor symmetry).
If this is true, then, the observed lepton mixing $U_{PMNS}$ will 
also deviate from the exact tribimaximal mixing $U_{PMNS}=U_{TB}$
by additional effects which gives the deviation from 
$V_{CKM} ={\bf 1}$. 
(Also see, e.g., Ref.\cite{Tanimoto02} for a possible 
deviation of $U_{PMNS}$ from a bimaximal mixing (not 
tribimaximal mixing) related to $V_{CKM}$.)

Recently, Datta \cite{Datta} has investigated possible flavor 
changing neutral current processes using the same assumption
that $V_{CKM}={\bf 1}$ and $U_{PMNS}=U_{TB}$ in a flavor symmetry limit.
By using a specific mass matrix model, he have discussed realistic 
mixings $V_{CKM}$ and $U_{PMNS}$ caused by a small breaking of 
the flavor symmetry.  
Also, Plentinger and Rodejohann \cite{Rode05} have 
investigated possible deviations from tribimaximal mixing by 
assuming a special form of the neutrino mass matrix.
Furthermore, there are many works which discuss specific mass 
matrix models from the point of the so-called 
``quark-lepton-complementarity" \cite{QLC}.
In this paper, however, we start only from putting a simple ansatz 
stated later (in Eqs.(9) and (10)), without referring to any mass 
matrix model explicitly.

For convenience of later discussions, we define 
the tribimaximal mixing by a form
$$
U_{TB} = P_L^* U_{TB}^0 P_R ,
\eqno(2)
$$
where
$$
\begin{array}{l}
P_L ={\rm diag}(e^{i \alpha_1}, e^{i \alpha_2}, e^{i \alpha_3}) , \\
P_R ={\rm diag}(e^{i \gamma_1}, e^{i \gamma_2}, e^{i \gamma_3}) ,
\end{array}
\eqno(3)
$$
by including freedom of the phase convention, 
although the tribimaximal mixing is conventionally expressed by 
the form (1).
The purpose of the present paper is to speculate a possible
form of the lepton mixing matrix $U_{PMNS}$ under the ansatz 
$V_{CKM}={\bf 1} \leftrightarrow U_{PMNS}=U_{TB}$. 
We show, as stated later, that a natural realization of this 
ansatz leads to a simple relation
$$
U_{PMNS}=V_{CKM}^\dagger U_{TB} .
\eqno(4)
$$
By using the observed CKM mixing parameters, we estimate 
values of the neutrino oscillation parameters 
$\sin^2 \theta_{13}$, $\tan^2 \theta_{12}$
and $\sin^2 2\theta_{23}$, which are defined by
$$
\begin{array}{l}
\sin^2\theta_{13} \equiv |(U_{PMNS})_{13}|^2 , \\ 
\tan^2 \theta_{12} \equiv |(U_{PMNS})_{12}/(U_{PMNS})_{11}|^2 , \\
\sin^2 2 \theta_{23} \equiv 4 |(U_{PMNS})_{23}|^2 |(U_{PMNS})_{33}|^2 . 
\end{array}
\eqno(5)
$$

First, let us give conventions 
of the mass matrices:
the quark and charged lepton mass matrices $M_f$ ($f=u,d,e$)
are defined by the mass terms $\overline{f}_L M_f f_R$, so that
those are diagonalized as
$$
U_{fL}^\dagger M_f U_{fR} = D_f \equiv 
{\rm diag}(m_{f1}, m_{f2}, m_{f3}) ,
\eqno(6)
$$
and the neutrino (Majorana) mass matrix $M_\nu$ is defined
by $\overline{\nu}_L M_\nu \nu_L^c$, so that it is diagonalized as
$$
U_{\nu L}^\dagger M_\nu U_{\nu L}^* = D_\nu \equiv 
{\rm diag}(m_{\nu 1}, m_{\nu 2}, m_{\nu 3}) .
\eqno(7)
$$
Therefore, the quark and lepton mixing matrices, $V_{CKM}$ and 
$U_{PMNS}$, are given by
$$
V_{CKM}=U_{u L}^\dagger U_{d L} , \ \ \ \ U_{PMNS}= U_{e L}^\dagger U_{\nu L},
\eqno(8)
$$
respectively.
Hereafter, we refer to a flavor basis on which the mass matrix
$M_f$ is diagonal (i.e. $D_f$) as ``$f$-basis".
For example, in the $u$-basis, up-quark, down-quark, charged-lepton
and neutrino mass matrices are given by $D_u=U_{uL}^\dagger M_u U_{uR}$,
$M_d^{(u)}=U_{uL}^\dagger M_d U_{uR}$, 
$M_e^{(u)}=U_{uL}^\dagger M_e U_{uR}$ and
$M_\nu^{(u)}=U_{uL}^\dagger M_\nu U_{uL}^*$, respectively.

\vspace{3mm}

\noindent{\large\bf 2 \ Ansatz and speculation}

Let us mention an ansatz which leads to the relation (4).
We put the following ansatz: 
In the limit of $U_{dL} \rightarrow {\bf 1}$, 
the matrix $U_{eL}$ also becomes a unit matrix ${\bf 1}$, while
the matrix $U_\nu$ becomes the exact tribimaximal mixing $U_{TB}$ 
in the limit of $U_{uL} \rightarrow {\bf 1}$.
In other words, in the $u$-basis, the neutrino mass matrix 
$M_\nu^{(u)} \equiv U_{u L}^\dagger M_\nu U_{u L}^*$ is
diagonalized by the exact tribimaximal mixing matrix $U_{TB}$, i.e.
$$
U_{TB}^\dagger M_\nu^{(u)} U_{TB}^* = D_\nu .
\eqno(9)
$$ 
Here, we have supposed that, in a symmetry limit, i.e. when
an origin which causes $V_{CKM}\neq {\bf 1}$ is switched off,
the physical mass matrices $M_f$ become the diagonal forms
$D_f$, while the neutrino mass matrix $M_\nu$ becomes a specific
form $M_\nu^{(u)}$ defined by (9):
$$
(M_u,M_d;\, M_e, M_\nu) \rightarrow 
(D_u, D_d;\, D_e, U_{TB}D_\nu U_{TB}^T) .
\eqno(10)
$$
In other words, we consider that a common origin in the down sector
causes $D_d \rightarrow M_d$ and $D_e \rightarrow M_e$,
and a common origin in the up sector causes $D_u \rightarrow M_u$
and $U_{TB}D_\nu U_{TB}^T \rightarrow M_\nu$.
Of course, this transformation (10) can not be realized by a 
flavor-basis transformation, because $M_f$ and $D_f$ are
connected by Eqs.(6) and (7).
It is well-known that physics at a low-energy is unchanged 
under any flavor-basis transformation.

The ansatz (9) states that the mixing matrix $U_{\nu L}$ in the
neutrino sector, which is defined by 
$U_{\nu L}^\dagger M_\nu U_{\nu L}^* =D_\nu$, is given by
$$
U_{\nu L} = U_{uL} U_{TB} ,
\eqno(11)
$$
because $D_\nu =U_{TB}^\dagger M_\nu^{(u)} U_{TB}^* =U_{TB}^\dagger
(U_{uL}^\dagger M_\nu U_{uL}^*) U_{TB}^*$.
Therefore, 
the observed lepton mixing matrix $U_{PMNS}$ is given by
$$
U_{PMNS}= U_{eL}^\dagger U_{\nu L} = U_{eL}^\dagger U_{uL} U_{TB}
=U_{ed} V_{CKM}^\dagger  U_{TB} ,
\eqno(12)
$$
where $U_{ed}$ is a flavor-basis transformation matrix
defined by
$$
U_{ed}=U_{eL}^\dagger U_{d L} .
\eqno(13)
$$
(The relation (12) is also derived by using relations
$U_{\nu L}^{(u)} =U_{TB}$ and $U_{eL}^{(u)}=U_{uL}^\dagger
U_{eL}$ in the $u$-basis as $U_{PMNS}= U_{eL}^{(u) \dagger}
U_{\nu L}^{(u)} = U_{eL}^\dagger U_{uL} U_{TB}=U_{ed} 
V_{CKM}^\dagger  U_{TB}$.)
According to this notation, the CKM mixing matrix $V_{CKM}$
is expressed as $V_{CKM}=U_{ud}$.
Since $U_{ed}=U_{ue}^\dagger U_{ud} =U_{ue}^\dagger V_{CKM}$,
if we consider $U_{ue}={\bf 1}$, we obtain $U_{ed}=V_{CKM}$,
so that we will obtain $U_{PMNS} = U_{TB}$ from the  
relation (11).
However, such a case $U_{eu}={\bf 1}$ is unlikely under our
ansatz $U_{eL}\rightarrow {\bf 1}$ in the limit of 
$U_{dL} \rightarrow {\bf 1}$.
Generally speaking, $U_{ue}$ can vary from $U_{ue}={\bf 1}$ 
to $U_{ue}=V_{CKM}$, so that $U_{ed}$ varies from $U_{ed}=V_{CKM}$ 
to $U_{ed}={\bf 1}$ and Eq.(12) varies from $U_{PMNS}= U_{TB}$ to
$U_{PMNS}=V_{CKM}^\dagger U_{TB}$.
(Here, we have considered that $U_{ue}$ does, at least, not take 
a large mixing more than $V_{CKM}$ and a rotation to an opposite
direction, $V_{CKM}^\dagger$.)
Therefore, we can consider that the relation (4) describes
a maximal deviation of $U_{PMNS}$ from $U_{TB}$.
In spite of such a general consideration, 
we think that the case $U_{ed}={\bf 1}$ (or highly 
$U_{ed} \simeq {\bf 1}$) is a most natural realization of our ansatz
(10), because it means 
$U_{eL}\rightarrow {\bf 1}$ in the limit $U_{dL}\rightarrow {\bf 1}$.
Therefore, in this paper, we adopt the case $U_{ed}={\bf 1}$,
and investigate possible numerical values of the
neutrino oscillation parameters $\sin^2 \theta_{13}$, 
$\tan^2 \theta_{12}$ and $\sin^2 2\theta_{23}$ 
under the relation (4).

By the way, we are also interested in whether those values are dependent 
on the phase parameters $\alpha_i$ and $\gamma_i$ defined in Eq.(3).
The relation (12) is invariant under the rephasing
$U_{fL} \rightarrow U_{fL} P_f$ ($f=u,d,e$)
because of $V_{CKM} \rightarrow P_u^* V_{CKM} P_d$,
$U_{PMNS}\rightarrow P_e^* U_{PMNS}$, $U_{ed} \rightarrow P_e^* U_{ed} P_d$
and $U_{TB}\rightarrow P_u^* U_{TB}$ under the rephasing
(note that $U_{\nu L}$ does not have such a freedom of
rephasing).
Therefore, the phase matrices $P_L$ and $P_R$ originate in the 
mass matrix $M_\nu^{(u)}$ as shown in Eq.(9).
Then, Eq.(9) can be rewritten as
$$
(U_{TB}^0)^T \widetilde{M}_\nu^{(u)} U_{TB}^0 = D_\nu P_R^2 , 
\eqno(14)
$$
where
$$
\widetilde{M}_\nu^{(u)} = P_L M_\nu^{(u)} P_L . 
\eqno(15)
$$
Since the matrix $U_{TB}^0$ is orthogonal, the
mass matrix $\widetilde{M}_\nu^{(u)}$ has to be real.
In other words, the phase matrix $P_L$ is determined from 
the form $M_\nu^{(u)}$ so that $\widetilde{M}_\nu^{(u)}$ is real.
On the other hand, the phase matrix $P_R$ is fixed
so that $D_\nu P_R^2$ is real.
Then, we find that the numerical results for $|(U_{PMNS})_{ij}|$
are independent of the phases $\gamma_i$ in $P_R$, because 
$U_{PMNS}$ is expressed by 
$U_{PMNS}=U_{PMNS}^{P_R={\bf 1}} P_R$, 
so that the quantities 
$|(U_{PMNS})_{ij} |= |(U_{PMNS}^{P_R={\bf 1}})_{ij} e^{i\gamma_j}|$
are independent of the phase parameters $\gamma_j$.
The results are only dependent on the phase parameters
$\alpha_i$ in $P_L$.
Hereafter, for simplicity, we put $P_R={\bf 1}$.

Let us show that the neutrino oscillation parameters
$\sin^2\theta_{13}$ and $\sin^2 2\theta_{23}$ are only dependent 
on a relative phase parameter $\alpha\equiv\alpha_3-\alpha_2$.
Since $(U_{PMNS})_{i3}$ is expressed as
$$
(U_{PMNS})_{i3} =\sum_k (V_{CKM})_{ki}^* e^{-i \alpha_k} (U_{TB}^0)_{k3} 
= \frac{1}{\sqrt2} \left[ - (V_{CKM})_{2i}^* e^{- i\alpha_2}
+ (V_{CKM})_{3i}^* e^{- i\alpha_3} \right],
\eqno(16)
$$
the values $|(U_{PMNS})_{i3}|$ are dependent only on the 
parameter $\alpha$.
We illustrate the behaviors of $\sin^2\theta_{13}$ 
and $\sin^2 2\theta_{23}$ versus  
$\alpha$ in Fig.1 and Fig.2,  respectively.
Here, for numerical evaluation, we have used the Wolfenstein 
parameterization \cite{Wolfen} of $V_{CKM}$ and the best-fit values
\cite{PDG06}  $\lambda =0.2272$, $A=0.818$, $\overline{\rho}=0.221$
and $\overline{\eta}=0.340$.
We find that the values $\sin^2\theta_{13}$ and 
$\sin^2 2\theta_{23}$ are almost insensitive to the value
$\alpha$, and those take $\sin^2\theta_{13}=0.024-0.028$
and $\sin^2 2\theta_{23}=0.94-0.95$.
Those values are consistent with the present experimental data.
As shown in Fig.1, if we take the result  
$\sin^2\theta_{13}=0.016\pm 0.010$  ($1\sigma$) obtained 
from a global analysis of neutrino oscillation data 
by Fogli {\it et al.} \cite{Fogli08}, we can obtain allowed 
bounds for $\alpha$.
The sizable value  $\sin^2 \theta_{13}$ is within
a reach of forthcoming neutrino experiments planning by Double 
Chooz, Daya Bay, RENO, OPERA, and so on.
The value $\sin^2 2\theta_{23}=0.94-0.95$ is consistent with
the present observed value \cite{MINOS} 
$\sin^2 2\theta_{23}=1.00_{-0.13}$,
and the predicted value will also be testable soon by precision 
measurements in solar and reactor neutrino experiments.

Previously, Plentinger and Rodejohann \cite{Rode05} have predicted 
possible deviations from tribimaximal mixing by assuming a specific 
form of the neutrino mass matrix and by assuming a CKM-like hierarchy
of the mixing angles ($\theta_{12}^e=\lambda$, 
$\theta_{23}^e= A\lambda^2$, $\theta_{13}^e= B\lambda^3$) in 
the charged lepton sector.
Furthermore, they have assumed the quark-lepton-complementarity (QLC)
\cite{QLC}, and put an ad hoc relation $\theta_{12}^e=\theta_C$ 
($\theta_C$ is the Cabibbo mixing angle). 
Then, they have obtained a relation
$$
|(U_{PMNS})_{13}| \simeq \frac{1}{\sqrt2} |(V_{CKM})_{us}| .
\eqno(17)
$$
Their result (17) agrees with our result 
$\sin^2 \theta_{13}=0.024-0.028$, because
$$
|U(_{MNS})_{13}|^2=\frac{1}{2}\left|(V_{CLM})_{cd}^*-(V_{CKM})_{td}^*
e^{-i\alpha}\right|^2 \simeq \frac{1}{2} |(V_{CKM})_{us}|^2 \simeq 0.025,
\eqno(18)
$$
from Eq.(16). 

On the other hand, for the value $\tan^2 \theta_{12}$,
there is no simple situation (one-parameter dependency).
The values $(U_{PMNS})_{11}$ and $(U_{PMNS})_{12}$ are given by
$$
(U_{PMNS})_{11} = \frac{1}{\sqrt6} \left[ 2(V_{CKM})_{11}^* e^{- i\alpha_1}
-(V_{CKM})_{21}^* e^{- i\alpha_2} - (V_{CKM})_{31}^* e^{- i\alpha_3} \right],
\eqno(19)
$$
$$
(U_{PMNS})_{12} = \frac{1}{\sqrt3} \left[ (V_{CKM})_{11}^* e^{- i\alpha_1}
+(V_{CKM})_{21}^* e^{- i\alpha_2} + (V_{CKM})_{31}^* e^{- i\alpha_3} \right],
\eqno(20)
$$
so that the values $|(U_{PMNS})_{11}|$ and $|(U_{PMNS})_{12}|$ depend 
not only on $\beta\equiv \alpha_2-\alpha_1$ but also on
$\alpha \equiv \alpha_3-\alpha_2$.
However, since the observed CKM matrix parameters show
$1 \gg |(V_{CKM})_{cd}|^2 \gg |(V_{CKM})_{td}|^2$, we can neglect the
terms $(V_{CKM})_{31}^* e^{-i\alpha_3} $ compared with
$(V_{CKM})_{11}^* e^{-i\alpha_1} $ and $(V_{CKM})_{21}^* e^{-i\alpha_2}$,
so that the value $\tan^2 \theta_{12}$ approximately  
depends on only the parameter $\beta$.
We illustrate the behavior of $\tan^2\theta_{12}$
versus $\beta\equiv \alpha_2-\alpha_1$ in Fig.3, 
in which we take typical values of $\alpha$ such as  
$\alpha=0$ and $\alpha =-2\pi/3$.
We can see that $\tan^2\theta_{12}$ is, in fact, insensitive 
to the parameter $\alpha$.
In contrast to the cases of $\sin^2\theta_{13}$ 
and $\sin^2 2\theta_{23}$, the value of $\tan^2\theta_{12}$
are highly sensitive to the parameter $\beta$ as shown by
$$
|(U_{PMNS})_{12}|\simeq \frac{1}{\sqrt3} \left[ 1-|(V_{CKM})_{us}| 
\cos\beta \right] ,
\eqno(21)
$$
from Eq.(20).
The similar result has been obtained by Plentinger and Rodejohann 
\cite{Rode05}.
The value of $\tan^2\theta_{12}$ takes from 0.24 to 1.00 according 
to the variation 
in  $\beta$.
In order to fit the observed value \cite{KamLAND}
$\tan^2\theta_{12}\simeq 0.5$,
we must take $\beta \simeq \pm \pi/2$.
This will put a constraint on scenarios which give a
tribimaximal mixing.

Note that, from the relation (4), we can obtain a $CP$ violating
observable
$$
J_{CP}^\nu \simeq -\frac{1}{6} |(V_{CKM})_{us}| \sin\beta ,
\eqno(22)
$$
as well as in a model given in Ref.\cite{Rode05}.
Therefore, if we require a maximal $CP$ violation in the lepton
sector, we obtain $\beta \simeq \pm \pi/2$ as pointed out in 
Ref.\cite{Rode05}, which is compatible with the constraint from 
the observed value $\tan^2 \theta_{12} \simeq 0.5$.\cite{Rode08}

\vspace{3mm}

\noindent{\large\bf 3 \ Summary}

In conclusion, under the ansatz 
``$U_{PMNS} \rightarrow  U_{TB}$ 
in the limit of $V_{CKM} \rightarrow {\bf 1}$", 
we have 
speculated a simple relation $U_{PMNS}=V_{CKM}^\dagger U_{TB}$.
We have not referred an explicit mechanism (model) which gives
such a CKM mixing $V_{CKM}={\bf 1}$ in the limit of $U_{PMNS}=U_{TB}$.
For example, a model \cite{Rode05} by Plentinger and Rodejohann
is one of mass matrix models which explicitly realize our ansatz
because they have put an ad hoc assumption 
$\sin \theta_{12}^e = \sin\theta_{C}$.
A model \cite{Datta} by Datta is also one of such models.
However, such a model-building is not a purpose of the present paper. 
We have started our investigation by admitting the relation
$U_{PMNS}\rightarrow U_{TB}$ as $V_{CKM}\rightarrow {\bf 1}$
as an ansatz.
The relation $U_{PMNS}=V_{CKM}^\dagger U_{TB}$ is widely valid 
for all models which are consistent with our ansatz.

By using the observed CKM matrix parameters, we have estimated
the lepton mixing parameters $\sin^2\theta_{13}$, 
$\sin^2 2\theta_{23}$ and $\tan^2 \theta_{12}$.
The values of $\sin^2 2\theta_{23}$ and $\sin^2 \theta_{13}$ 
are almost independent of the phase convention, and they 
take values $\sin^2 \theta_{13} = 0.024-0.028$ and
$\sin^2 2\theta_{23} = 0.94-0.95$. 
The sizable value of $\sin^2 \theta_{13}$ is within
a reach of forthcoming neutrino experiments planning by Double 
Chooz, Daya Bay, RENO, OPERA, and so on.
The value of $\sin^2 2\theta_{23}$ is also testable soon 
by precision measurements in solar and reactor 
neutrino experiments.
On the other hand, the value of $\tan^2 \theta_{12}$ has highly 
depended on the phase convention of the tribimaximal mixing, 
and the value has been in a range $0.24<\tan^2 \theta_{12}<1.00$.
Note that the phase matrix $P_L$ cannot be absorbed  into the 
rephasing of $V_{CKM}$, although it seems to be possible from 
the expression (4).
Since the present observed value of $\tan^2 \theta_{12}$ is
$\tan^2 \theta_{12} \simeq 0.5$, the phase parameter
$\beta$ is constrained as 
$\beta \simeq \pm \pi/2$.
This put a strong constraint on models which lead to the exact
tribimaximal mixing (2).  
The requirement of a maximal $CP$ violation in the lepton
sector is interestingly related to the observed value 
$\tan^2 \theta_{12} \simeq 0.5$.

If the predicted values $\sin^2 \theta_{13}= 0.024-0.028$ and 
$\sin^2 2\theta_{23}\simeq 0.94-0.95$ are denied by forthcoming
neutrino oscillation experiments, it means a denial of the 
simple view that the lepton mixing $U_{PMNS}$ becomes 
the exact tribimaximal mixing $U_{TB}$ in the limit of 
$V_{CKM} \rightarrow {\bf 1}$.
We will be compelled to consider that the view stated above
is oversimplified and the situation of quark and lepton
flavor mixings is more complicated.
The observed values of neutrino oscillation parameters will 
provide us a promising clue to a possible structure of 
$U_{ed}$, although we simply assumed $U_{ed}={\bf 1}$ 
in the expression (12). 
This will shortly become clear by forthcoming experiments.

\vspace{3mm}

\centerline{\large\bf Acknowledgments} 

The authors would like to thank T.~Kubota, M.~Tanimoto,
A.~Datta and W.~Rodejohann 
for giving useful information on references. 
The authors also thank W.~Rodejohann for his comment on the
$CP$ violating phase $\beta$.
One of the authors (YK) is supported by the Grant-in-Aid for
Scientific Research, Ministry of Education, Science and 
Culture, Japan (No.18540284).

\vspace{4mm}

\newpage

{\scalebox{0.9}{\includegraphics{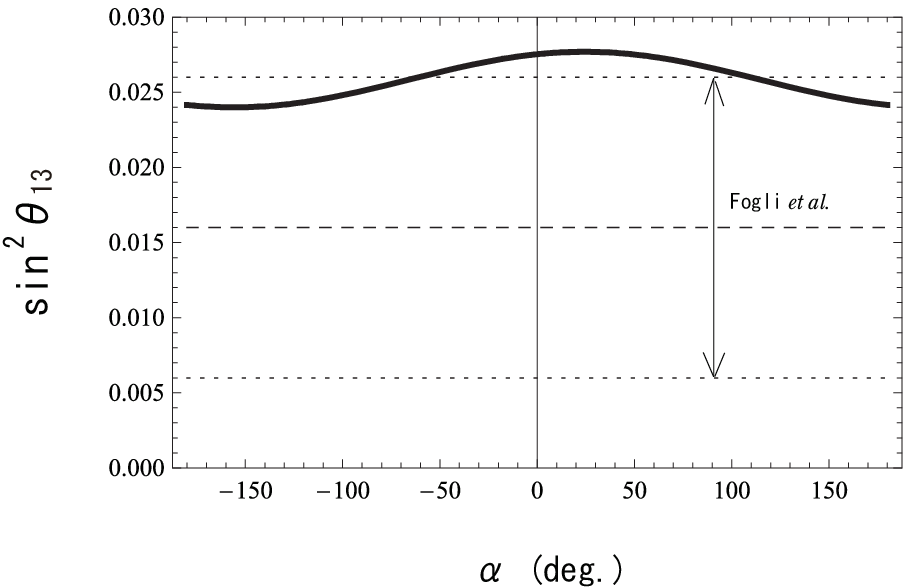}} }

\begin{quotation}
{\bf Fig.~1}  Behavior of $\sin^2 \theta_{13}$
versus $\alpha=\alpha_3-\alpha_2$. 
The horizontal dashed and dotted lines denote
the analysis $\sin^2 \theta_{13} = 0.016\pm 0.010$ ($1\sigma$)
by Fogli {\it et al.} \cite{Fogli08}.
\end{quotation}

\vspace{5mm}

{\scalebox{0.9}{\includegraphics{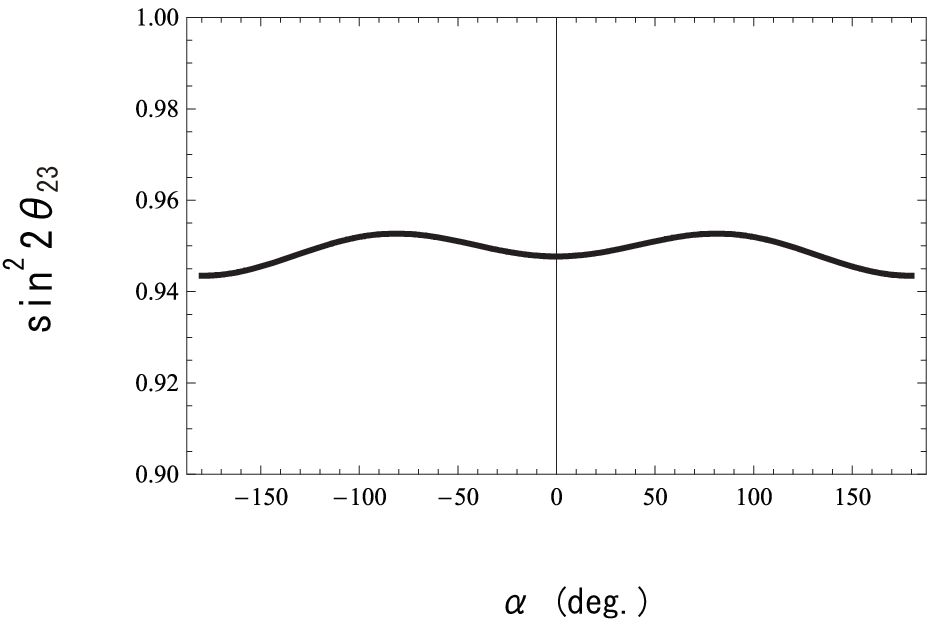}} }

\begin{quotation}
{\bf Fig.~2}  Behavior of $\sin^2 2\theta_{23}$
versus $\alpha=\alpha_3-\alpha_2$. 
The predicted value is consistent with the 
observed data \cite{MINOS} $\sin^2 2\theta_{23}=1.00_{-0.13}$.

\end{quotation}

\vspace{5mm}

{\scalebox{0.9}{\includegraphics{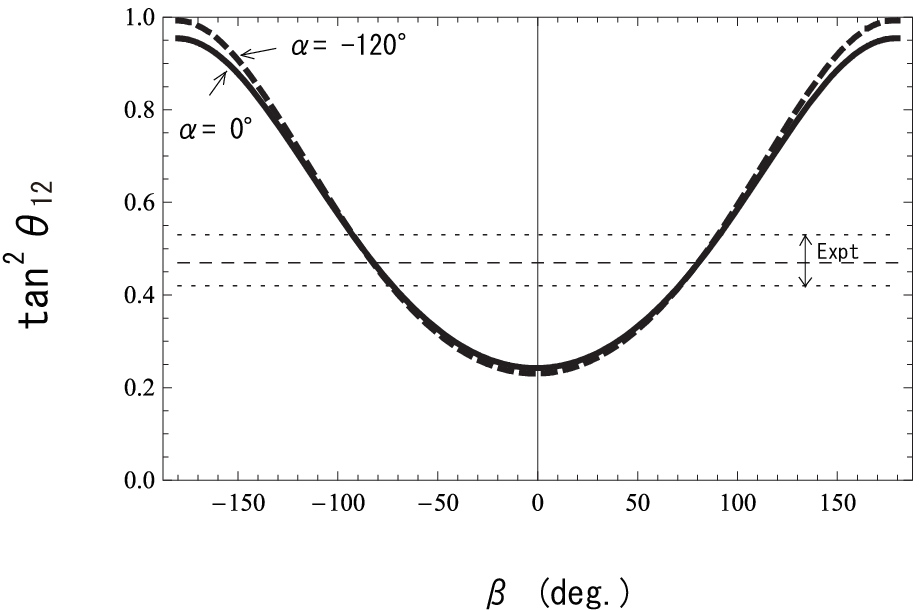}} }

\begin{quotation}
{\bf Fig.~3}  Behavior of $\tan^2 \theta_{12}$
versus $\beta=\alpha_2-\alpha_1$. 
The horizontal dashed and dotted lines denote the observed
values \cite{KamLAND} $\tan^2\theta_{12}=0.47^{+0.06}_{-0.05}$.
\end{quotation}

\vspace{10mm}


\begin{thebibliography}{99}
%
\bibitem{U13}
A.~S.~Joshipura, arXiv: hep-ph/0411154; 
C.~H.~Albright and M.~C.~Chen, Phys.~Rev. {\bf D74} (2006) 113006.
%
\bibitem{tribi} 
P.~F.~Harrison, D.~H.~Perkins and W.~G.~Scott,
 Phys.~Lett. {\bf B458} (1999) 79;
 Phys.~Lett. {\bf B530} (2002) 167;
Z.-z.~Xing, Phys.~Lett. {\bf B533} (2002) 85;
P.~F.~Harrison and W.~G.~Scott,  Phys.~Lett. {\bf B535}  (2003) 163;
Phys.~Lett. {\bf B557} (2003) 76;
E.~Ma, Phys.~Rev.~Lett. {\bf 90} (2003) 221802;
C.~I.~Low and R.~R.~Volkas, Phys.~Rev. {\bf D68} (2003) 033007.
%
\bibitem{KamLAND} S.~Abe, {\it et al.}, KamLAND collaboration,
Phys.~Rev.~Lett. {\bf 100} (2008) 221803.
%
\bibitem{MINOS} D.~G.~Michael {\it et al.}, MINOS collaboration,
Phys.~Rev.~Lett. {\bf 97} (2006)  191801;
J.~Hosaka, {\it et al.}, Super-Kamiokande collaboration, Phys.~Rev. 
{\bf D74} (2006) 032002.
%
\bibitem{Fogli08}
G.~Fogli, E.~Lisi, A.~Marrone, A.~Palazzo and A.~M.~Rotunno,
arXiv:0806.2649 [hep-ph].
%
%
\bibitem{MNS}
Z.~Maki, M.~Nakagawa and S.~Sakata, Prog.~Theor.~Phys. 
{\bf 28} (1962) 870;
B.~Pontecorvo, Zh.~Eksp.~Theor.~Fiz. {\bf 53} (1967) 1717;
Sov.~Phys. JETP {\bf 26} (1968) 984.
%
\bibitem{CKM}
N.~Cabibbo, Phys.~Rev.~Lett. {\bf 10} (1963) 531;
M.~Kobayashi and T.~Maskawa, Prog.~Theor.~Phys. {\bf 49}
(1973) 652.
%
\bibitem{Tanimoto02}
C.~Giunti and M.~Tanimoto, Phys.~Rev. {\bf D66} (2002) 053013.
%
\bibitem{Datta}
A.~Datta, arXiv:0807.0795 [hep-ph].
%
\bibitem{Rode05}
F.~Plentinger and W.~Rodejohann, Phys.~Lett. {\bf B625} (2005) 264.
%
\bibitem{QLC}
M.~Raidal, Phys.~Rev.~Lett. {\bf 93} (2004) 161801;
H.~Minakata and A.~Y.~Smirnov, Phys.~Rev. {\bf D70} (2004) 073009;
P.~H.~Frampton and R.~N.~Mohapatra, JHEP {\bf 0501} (2005) 025;
J.~Ferrandis and S.~Pakvasa, Phys.~Rev. {\bf D71} (2005) 033004;
S.~Antush, S.~F.~King and R.~N.~Mohapatra, Phys.~Lett. 
{\bf B618} (2005) 150;
S.~K.~Kang, C.~S.~Kim and J.~Lee, Phys.~Lett. {\bf B619} (2005) 129;
K.~Cheung {\it et al.}, Phys.~Rev. {\bf D72} (2005) 036003;
A.~Datta, L.~Everett and P.~Ramond, Phys.~Lett. {\bf B620} (2005) 42.
For recent developments, see, e.g. 
F.~Plentinger and G.~Seidl, Phys.~Rev. {\bf D78} (2008) 045004. 
%
\bibitem{Wolfen} L.~Wolfenstein, Phys.~Rev.~Lett. {\bf 51} (1983) 1945.
%
\bibitem{PDG06} Particle Data Group, {J.~Phys.~G} {\bf 33} (2006) 1.
%
\bibitem{Rode08} W.~Rodejohann, private communication.
%
\end{thebibliography}
\end{document}